\documentclass[final,5p,times,twocolumn]{elsarticle}
\usepackage{amsmath}
\usepackage{bm}
\usepackage{booktabs}
\usepackage{braket}
\usepackage{caption}
\usepackage{color}
\usepackage{graphicx}
\usepackage[colorlinks]{hyperref}
\usepackage{multirow}

\journal{Journal of Subatomic Particles and Cosmology}

\begin{document}

\begin{frontmatter}
  \title{Sound velocity in quarkyonic matter within the quark-meson coupling model}
  \author[1,2]{Tsuyoshi Miyatsu}
  \address[1]{National Institute of Technology, Asahikawa College, Asahikawa 071-8142, Japan}
  \address[2]{Department of Physics and OMEG Institute, Soongsil University, Seoul 06978, Republic of Korea}
  \ead{miyatsu@asahikawa-nct.ac.jp, tsuyoshi.miyatsu@ssu.ac.kr}
  \author[2]{Myung-Ki Cheoun}
  \ead{cheoun@ssu.ac.kr}
  \author[3]{Koichi Saito}
  \address[3]{Department of Physics and Astronomy, Faculty of Science and Technology, Tokyo University of Science, Noda 278-8510, Japan}
  \ead{koichi.saito@rs.tus.ac.jp}

  \begin{abstract}
    We investigate the sound velocity in dense symmetric nuclear matter within the quarkyonic quark-meson coupling (QQMC) model, which combines quark-level Pauli blocking with the medium modification of nucleon structure.
    In the original formulation, the sound velocity diverges at the quark saturation density.
    Motivated by the possible smearing of the sharp Fermi surface in the nucleon momentum distribution, we introduce a phenomenological infrared regulator into the boundary condition.
    This prescription removes the divergence and the accompanying discontinuities in the pressure and nucleon chemical potential.
    We then compare the calculated pressure with constraints inferred from collective flow and kaon production data in heavy-ion collisions.
    The empirical pressure constraints favor a large width, leading to a gradual increase in the sound velocity above quark saturation rather than a pronounced peak near its onset.
  \end{abstract}


\end{frontmatter}

\section{Introduction}
\label{sec:introduction}

At densities well above the nuclear saturation density, $\rho_{0}$, not only nucleons, $N$, but also exotic degrees of freedom, such as hyperons, meson condensates, and deconfined quarks, may play important roles in determining the properties of dense matter~\citep{Glendenning:1991es,Schaffner:1995th,Glendenning:1998zx,Miyatsu:2013yta,Miyatsu:2025rzn,Miyatsu:2013hea,Muto:2021jms}.
In particular, how the transition from hadronic matter to quark matter occurs remains one of the central issues in nuclear physics and astrophysics~\citep{Itoh:1970uw,Masuda:2012kf,Miyatsu:2015kwa,Miyatsu:2017teh}.

Among various theoretical approaches to describing dense matter, the concept of quarkyonic matter, which was originally proposed in the large-$N_c$ limit of quantum chromodynamics (QCD), has recently attracted particular attention~\citep{McLerran:2007qj,Fukushima:2020cmk,Kojo:2021ugu,Kojo:2025vcq}.
In the quarkyonic picture, quarks or baryons exist in a dual manner, and Pauli blocking at the quark level leads to a considerable stiffening of the equation of state (EoS) and an enhancement of the sound velocity~\citep{Fujimoto:2023mzy,Koch:2024qnz,Ivanytskyi:2025cnn}.

In our previous study~\citep{Miyatsu:2026kmb}, we constructed the quarkyonic quark-meson coupling (QQMC) model by combining the dual quarkyonic picture with the quark-meson coupling (QMC) model~\citep{Saito:1994ki,Guichon:1995ue,Saito:2005rv}, in which a relativistic Gaussian quark wavefunction is adopted to describe the nucleon structure~\citep{Saito:2024qth}.
We found that the quark saturation density, $\rho_{\mathrm{sat}}$, strongly depends on the nucleon size and that the inclusion of nuclear interactions shifts the onset of quark saturation to lower density.
A large nucleon size leads to a narrower quark momentum distribution and enhances the occupation of low-momentum quark states, thereby lowering the quark saturation density.
Nevertheless, for the realistic nucleon sizes considered here, the relation $\rho_{\mathrm{sat}}>\rho_0$ holds in symmetric nuclear matter.

\begin{figure}[b!]
  \centering
  \includegraphics[width=\linewidth]{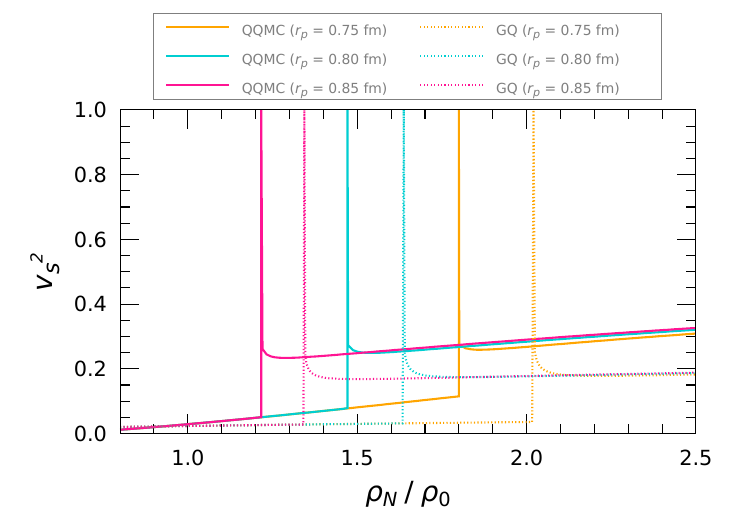}
  \caption{\label{fig:1}
    Squared sound velocity, $v_{s}^{2}$, as a function of baryon density ratio, $\rho_{N}/\rho_{0}$, in symmetric nuclear matter.
    The solid and dotted curves represent the unregularized QQMC and GQ models, respectively.
    We present three cases of the root-mean-square radius of the proton in vacuum, $r_{p}=0.75$, $0.80$, and $0.85$ fm, without an infrared regulator.
    The nuclear saturation density, $\rho_{0}$, is chosen to be $\rho_{0}=0.15$ fm$^{-3}$.}
\end{figure}

As shown in Fig.~\ref{fig:1}, however, the sound velocity diverges at $\rho_{\mathrm{sat}}$ in both the QQMC and noninteracting Gaussian quarkyonic (GQ) models.
Such singular behavior originates from the assumption of a sharp Fermi surface in the bulk part of the nucleon momentum distribution and should not be interpreted as a physical phenomenon.
Furthermore, the chemical potential and pressure are discontinuous at $\rho_{\mathrm{sat}}$.

In the present study, we focus on this issue and its implications for the EoS, following the formulation developed by \citet{Saito:2025yld}.
To remove such singular behavior, we introduce an infrared regulator into the boundary condition as a phenomenological prescription motivated by the possible smearing of the sharp Fermi surface.
We then examine its effects on the pressure, chemical potential, and sound velocity in symmetric nuclear matter.
In particular, by comparing the calculated pressure with constraints deduced from heavy-ion collision (HIC) data, we investigate
whether a pronounced peak in the sound velocity is an intrinsic feature of quarkyonic matter or depends sensitively on the treatment of the nucleon momentum distribution near quark saturation.

\section{Quarkyonic quark-meson coupling model}
\label{sec:QQMC}

We briefly explain the quarkyonic quark-meson coupling (QQMC) model developed in Refs.~\citep{Miyatsu:2026kmb,Saito:2025yld}.
To describe the quark structure of the nucleon, we adopt a relativistic confinement potential of the scalar-vector harmonic oscillator type, $U(r) = \frac{c}{2}(1+\gamma^0)r^2$, with the strength parameter $c$~\citep{Saito:2024qth}.
In the nuclear medium, the confined quarks interact with the mean fields of $\sigma$ and $\omega$ mesons, denoted as $\bar{\sigma}$ and $\bar{\omega}$.
The Dirac equation for the quark field is then given by
\begin{equation}
  \left[
    i\gamma\cdot\partial
    - (m-V_s) - \gamma^0 V_0
    - \frac{c}{2}(1+\gamma^0)r^2
  \right]\psi_Q(\bm{r},t)=0,
  \label{eq:Dirac}
\end{equation}
where $m$ is the constituent quark mass, and $V_s=g_\sigma^q\bar{\sigma}$ and $V_0=g_\omega^q\bar{\omega}$ with the quark-meson coupling constants, $g_\sigma^q$ and $g_\omega^q$.

Using the relativistic Gaussian quark wavefunction, we obtain the momentum distribution of a single quark in an in-medium nucleon,
\begin{equation}
  \varphi^{\ast}(\bm{q})
  = \pi^{3/2}a^{*3}
  \frac{16\lambda^{*2}a^{*2}}
  {2\lambda^{*2}a^{*2}+3}
  \left(1+\frac{q^2}{\lambda^{*2}}\right)
  e^{-a^{*2}q^2},
  \label{eq:quark_distribution}
\end{equation}
with $\lambda^{\ast}=\epsilon^{\ast}+m^{\ast}$ and $a^{*2}=1/\sqrt{c\lambda^{\ast}}$, where $m^{\ast}=m-V_s$ and $\epsilon^{\ast}=\epsilon-V_0$ are the effective quark mass and single-particle energy, respectively.
The effective nucleon mass, $M_N^{\ast}$, is calculated by taking into account the spin correlation and the center-of-mass correction, as explained in Ref.~\citep{Saito:2025yld}.

In the dual quarkyonic picture, the momentum-space distributions of quarks and nucleons, $f_Q(q)$ and $f_N(k)$, respectively, are related to each other through the sum rule
\begin{equation}
  f_Q(q)
  = \int_k
  \varphi^{\ast}\left(
    \bm{q}-\frac{\bm{k}}{N_c}
  \right) f_N(k),
  \label{eq:duality}
\end{equation}
where $\int_k \equiv \int d^3\bm{k}/(2\pi)^3$ and $N_c$ is the number of colors.
The single quark momentum distribution inside a nucleon is normalized as $\int_{q}\varphi^{\ast}(\bm{q})=1$.
Consequently, the momentum integrals of $f_{Q}$ and $f_{N}$ are identical, ensuring consistency between the quark and nucleon representations of baryon density.

Below the quark saturation density, $\rho_{\mathrm{sat}}$, the nucleon momentum distribution is given by the usual Fermi distribution, $f_N(k)=\theta(k_s^{\ast}-k)$.
At $\rho_{\mathrm{sat}}$, the quark momentum distribution reaches unity at $q=0$.
Above $\rho_{\mathrm{sat}}$, due to Pauli blocking at the quark level, the nucleon momentum distribution is assumed to be separated into two segments:
\begin{equation}
  f_N(k)
  = \frac{1}{N_c^3}\theta(k_b^{\ast}-k)
  + \theta(k_s^{\ast}-k)\theta(k-k_b^{\ast}),
  \label{eq:nucleon_distribution}
\end{equation}
where $k_b^{\ast}$ and $k_s^{\ast}$ denote the nucleon bulk and shell momenta, respectively.
The factor $1/N_{c}^{3}$ in the bulk region reflects the momentum-space scaling between quark and nucleon degrees of freedom and represents the suppression of low-momentum nucleon states due to quark-level Pauli blocking.
The two momenta are determined by the nuclear density,
\begin{equation}
  \rho_N
  = \frac{2}{3\pi^2}
  \left(k_s^{*3}-\beta k_b^{*3}\right),
  \label{eq:nuclear_density}
\end{equation}
and the boundary condition,
\begin{equation}
  f_Q(0,k_s^{\ast})-\beta f_Q(0,k_b^{\ast})=1,
  \label{eq:boundary}
\end{equation}
with $\beta=1-1/N_c^3$.
Here, $f_Q(q,k_i^{\ast})$ ($i=b,s$) denotes the quark momentum distribution obtained from Eq.~\eqref{eq:duality} with $f_N(k)=\theta(k_i^{\ast}-k)$.

In the QQMC model, the total energy density in symmetric nuclear matter is given by
\begin{equation}
  \varepsilon_{\rm tot}
  = \varepsilon_N
  + \frac{1}{2}\left(
    m_\sigma^2\bar{\sigma}^2
    + m_\omega^2\bar{\omega}^2
  \right)
  + \frac{1}{3}g_2\bar{\sigma}^3,
  \label{eq:energy_density}
\end{equation}
with
\begin{equation}
  \varepsilon_N
  = \frac{2}{\pi^2}\int dk\,k^2
  \sqrt{M_N^{*2}+k^2}\,f_N(k).
  \label{eq:nucleon_energy_density}
\end{equation}
Here, $g_{2}$ is the cubic self-coupling of the $\sigma$ meson.
The meson mean fields are calculated self-consistently at a given nuclear density, $\rho_{N}$.
For details of the formulation and the determination of the coupling constants, see Ref.~\citep{Saito:2025yld}.

Once the total energy density is obtained as a function of $\rho_N$, the chemical potential, $\mu_N$, pressure, $P$, and squared sound velocity, $v_{s}^{2}$, are calculated through the thermodynamic relations
\begin{align}
  \mu_N &= \frac{\partial\varepsilon_{\rm tot}}
          {\partial\rho_N}, \\
  P &= \rho_N\mu_N-\varepsilon_{\rm tot}, \\
  v_s^2 &= \frac{\partial P}
          {\partial\varepsilon_{\rm tot}}
          = \frac{\rho_N}{\mu_N}
          \frac{\partial\mu_N}{\partial\rho_N}.
          \label{eq:sound_velocity}
\end{align}

\section{Minimal corrections to the QQMC model}
\label{sec:corrections}

\begin{figure}[t!]
  \centering
  \includegraphics[width=\linewidth]{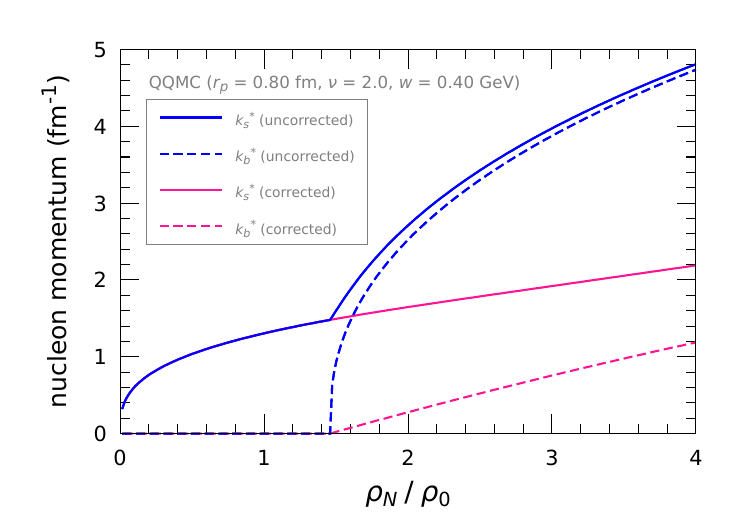}
  \caption{\label{fig:2}
    Density dependence of nucleon bulk and shell momenta, $k_{b}^{\ast}$ and $k_{s}^{\ast}$, in symmetric nuclear matter.
    The blue and magenta curves show the results without and with the regulator ($w=0.4$~GeV), respectively.}
\end{figure}

As shown in Fig.~\ref{fig:2}, the nucleon bulk and shell momenta, $k_b^{\ast}$ and $k_s^{\ast}$, increase rapidly just above $\rho_{\rm sat}$ in the uncorrected case.
This rapid increase leads to discontinuities in $\mu_{N}$ and $P$, and a divergence in $v^{2}_{s}$ at $\rho_{\rm sat}$.

The origin of such singular behavior can be understood from the boundary condition in Eq.~\eqref{eq:boundary}.
The first and second derivatives of $k_b^{\ast}$ with respect to $k_s^{\ast}$ behave near $\rho_{\rm sat}$ as
\begin{equation}
  \frac{\partial k_b^{\ast}}{\partial k_s^{\ast}}
  \propto (k_b^{\ast})^{-2}, \qquad
  \frac{\partial^2 k_b^{\ast}}{\partial k_s^{\ast 2}}
  \propto (k_b^{\ast})^{-5},
  \label{eq:singular_derivatives}
\end{equation}
because $f_Q(0,k_b^{\ast}) \propto k_b^{\ast3}$ as $k_b^{\ast} \to 0$.

To remove such singular behavior, we consider minimal corrections to the QQMC model.
Motivated by the possible smearing of the sharp Fermi surface in the nucleon momentum distribution, we modify the boundary condition that determines $k_b^{\ast}$ and $k_s^{\ast}$.
Such smearing should in principle arise from quark and gluon exchanges among neighboring nucleons, but it is extremely difficult to treat these effects explicitly in a many-body system.
Following Ref.~\citep{Saito:2025yld}, we thus introduce an infrared regulator into the boundary condition:
\begin{equation}
  f_Q(0,k_s^{\ast})
  - \beta \alpha_\nu(k_b^{\ast}) f_Q(0,k_b^{\ast})
  = 1,
  \label{eq:modified_boundary}
\end{equation}
with
\begin{equation}
  \alpha_\nu(k_b^{\ast})
  = 1 + \left(\frac{w}{k_b^{\ast}}\right)^\nu,
  \label{eq:regulator}
\end{equation}
where $w$ is a width parameter and $\nu$ is a positive real number.
Note that the original boundary condition is recovered when $w=0$.

For $w>0$, the modified boundary condition gives the following power laws as $k_b^{\ast}\to 0^{+}$,
\begin{equation}
  \frac{\partial k_b^{\ast}}{\partial k_s^{\ast}}
  \propto (k_b^{\ast})^{\nu-2},
  \qquad
  \frac{\partial^2 k_b^{\ast}}{\partial (k_s^{\ast})^2}
  \propto (k_b^{\ast})^{2\nu-5}.
\end{equation}
For $3/2<\nu<3$, the chemical potential and pressure are smooth and the sound velocity is continuous at $\rho_{\rm sat}$~\citep{Saito:2025yld}.

\section{Numerical results}
\label{sec:results}

In the present calculation, we take the nucleon size parameter to be $r_p=0.80$~fm, which gives $\rho_{\mathrm{sat}}\simeq 1.46\,\rho_0$ in the QQMC model (see Fig.~\ref{fig:1}).
To examine the effects of the minimal corrections, we fix $\nu=2.0$ and compare the results for $w=0.2$, $0.3$, and $0.4$~GeV.

Figure~\ref{fig:2} shows the nucleon bulk and shell momenta, $k_b^{\ast}$ and $k_s^{\ast}$, as functions of the nuclear density, $\rho_N$.
Without the regulator, both momenta increase rapidly just above $\rho_{\rm sat}$.
As $k_b^{\ast}$ approaches $k_s^{\ast}$, the occupied nucleon shell becomes thin.
With the regulator ($w=0.4$~GeV), both momenta increase more gradually, and the shell remains wider above $\rho_{\rm sat}$.

\begin{figure}[t!]
  \centering
  \includegraphics[width=\linewidth]{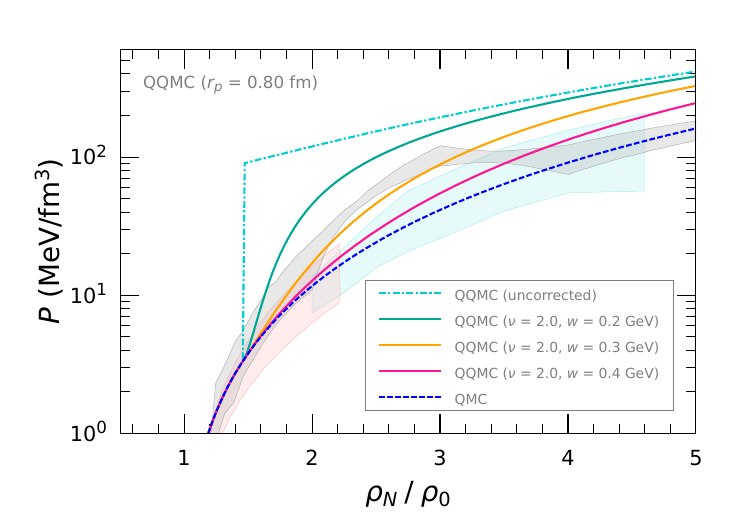}
  \caption{\label{fig:3}
    Pressure $P$ as a function of $\rho_N/\rho_0$ in symmetric nuclear matter.
    The solid curves show the corrected QQMC results for the three values of $w$.
    The dash-dotted and dashed curves show the uncorrected QQMC and conventional QMC results, respectively.
    The shaded areas indicate the empirical constraints~\citep{Danielewicz:2002pu,Oliinychenko:2022uvy,Fuchs:2005zg,Lynch:2009vc}.}
\end{figure}

Figure~\ref{fig:3} shows the pressure, $P$, together with the empirical constraints inferred from flow and kaon production data in HICs~\citep{Danielewicz:2002pu,Oliinychenko:2022uvy,Fuchs:2005zg,Lynch:2009vc}.
We also show the conventional QMC result for comparison.
In the uncorrected QQMC model, the pressure jumps at $\rho_{\rm sat}$.
The minimal corrections remove this discontinuity for all three values of $w$.
Above $\rho_{\rm sat}$, increasing $w$ lowers $P$.
Among the three values considered, $w=0.4$~GeV gives the closest overall agreement with the empirical pressure bands.
It should be noted that the empirical pressure bands are extracted from HIC data through transport-model analyses.
The preferred value of $w$ should therefore be understood within the present QQMC parametrization rather than as a model-independent constraint.

\begin{figure}[t!]
  \centering
  \includegraphics[width=\linewidth]{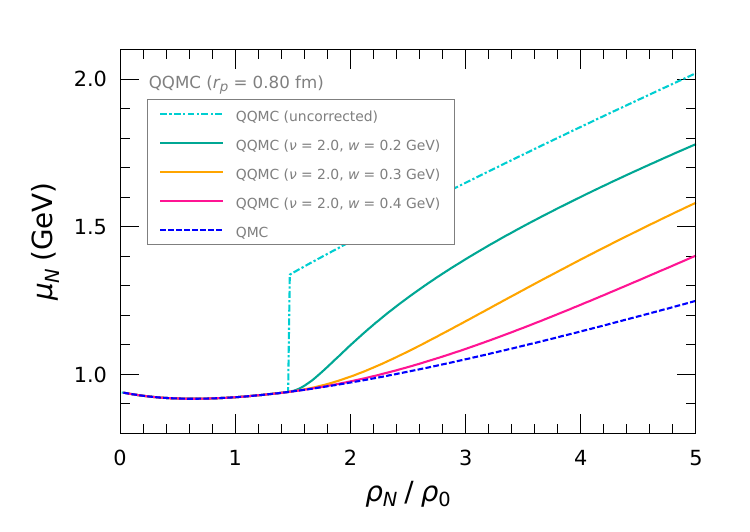}
  \caption{\label{fig:4}
    Nucleon chemical potential $\mu_N$ as a function of $\rho_N/\rho_0$.}
\end{figure}

As shown in Fig.~\ref{fig:4}, the nucleon chemical potential, $\mu_N$, also becomes continuous at $\rho_{\rm sat}$ with the minimal corrections.
Increasing $w$ makes the rise in $\mu_N$ more gradual just above $\rho_{\rm sat}$.
At high density, $\mu_N$ remains above the conventional QMC result for all three values of $w$.

\begin{figure}[t!]
  \centering
  \includegraphics[width=\linewidth]{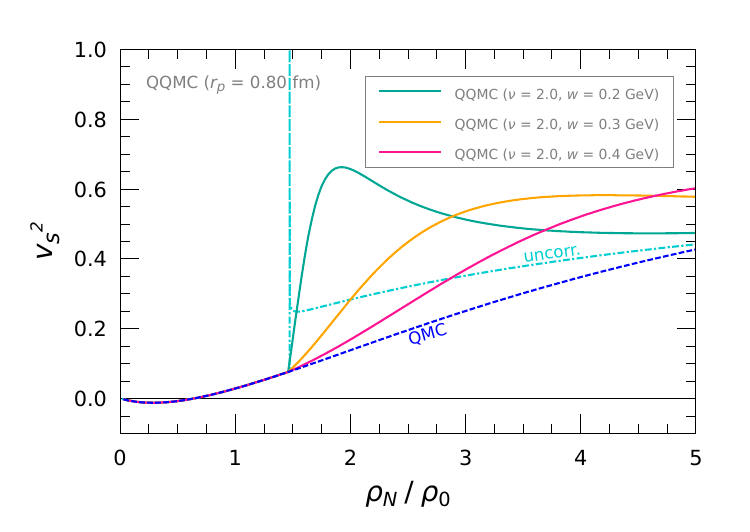}
  \caption{\label{fig:5}
    Squared sound velocity $v_s^2$ as a function of $\rho_N/\rho_0$.
    The solid curves show the corrected QQMC results for $w=0.2$, $0.3$, and $0.4$~GeV.
    The dash-dotted and dashed curves show the uncorrected QQMC and conventional QMC results, respectively.}
\end{figure}

Finally, Fig.~\ref{fig:5} shows the squared sound velocity, $v_s^2$.
With the regulator, $v_s^2$ is continuous at $\rho_{\rm sat}$ and stays below the causal limit throughout the quarkyonic regime shown.
For $w=0.2$~GeV, $v_s^2$ rises sharply to a peak of about 0.66 at $\rho_N\simeq 1.9\,\rho_0$ and then decreases.
For $w=0.3$~GeV, the peak is broader and shifts to higher density. For $w=0.4$~GeV, $v_s^2$ increases monotonically up to $5\,\rho_0$, where it reaches about 0.60.
Thus, within the QQMC model, the width favored by the empirical pressure bands gives a gradual rise in $v_s^2$ above quark saturation.
Even for this width, Pauli blocking at the quark level enhances $v_s^2$ relative to the conventional QMC result.

\section{Summary and conclusion}
\label{sec:summary}

We have examined the effects of minimal corrections to the QQMC model on the pressure, nucleon chemical potential, and sound velocity in symmetric nuclear matter.
Motivated by the smearing of the sharp Fermi surface in the nucleon momentum distribution, we have introduced an infrared regulator into the boundary condition.
We have shown that the regulator slows the growth of the nucleon bulk and shell momenta just above the quark saturation density.
The pressure and chemical potential then vary continuously through quark saturation, and the divergence in the sound velocity disappears.

We have found that the squared sound velocity, $v_s^2$, develops a pronounced peak above quark saturation for a small regulator width.
For $w=0.2$~GeV, $v_s^2$ peaks at about 0.66 near $1.9\,\rho_0$.
Increasing $w$ to 0.3~GeV lowers and broadens the peak and shifts it to higher density. We have compared the calculated pressure with the empirical bands inferred from heavy-ion collision data and found the closest overall agreement for $w=0.4$~GeV among the three widths examined.
For this width, $v_s^2$ rises monotonically to about 0.60 at $5\,\rho_0$. Thus, within the QQMC model, the pressure data disfavor a pronounced peak of $v_{s}^{2}$ just above quark saturation.

The regulator remains phenomenological, and a microscopic description of the smearing of the Fermi surface in terms of quark and gluon exchanges between neighboring nucleons is still needed.
A related question is whether the assumed boundary condition suffices to determine a quark momentum distribution above saturation that respects Fermi statistics and minimizes the energy.
The present regulator modifies the boundary condition without explicitly smearing the step functions in the nucleon momentum distribution.
It should therefore be regarded as an effective prescription for removing the thermodynamic singularities.
A fully microscopic treatment should determine the nucleon and quark momentum distributions consistently while satisfying the quark occupation constraint, $0 \leq f_{Q} \leq 1$.

We are considering extensions of the QQMC model to finite-temperature matter and to neutron-star matter in $\beta$ equilibrium.
\citet{Fujimoto:2024doc} have shown in a quarkyonic model that $d$-quark saturation can delay the onset of hyperons and suppress their occupation of low-momentum states.
Within the QQMC framework, we would like to examine how this mechanism affects the sound velocity and whether it can alleviate the hyperon puzzle by keeping the EoS sufficiently stiff to support neutron stars of about $2\,M_\odot$.



\section*{Data availability}

Data will be made available on request.



\section*{Acknowledgments}

This work was supported by the Basic Science Research Program through the National Research Foundation of Korea (NRF) under Grant Nos. RS-2026-25471436, RS-2025-16071941, RS-2023-00242196, and RS-2021-NR060129.

\bibliographystyle{elsarticle-num-names-mod.bst}
\bibliography{QCS2026.bib}

\end{document}